# Polarization Control and TM-Pass Filtering in SiN Photonics Integrated with 2D Multiferroic Materials

Ghada Dushaq, *Senior Member, IEEE,* Solomon Serunjogi, Srinivasa R. Tamalampudi, and Mahmoud Rasras, *Senior Member, IEEE*

*Abstract*— Polarization control is a critical function in integrated photonic circuits, directly impacting performance, stability, and signal integrity. In this work, we demonstrate the integration of multiferroic two-dimensional (2D) material CuCrP$_2$S$_6$ (CCPS) with silicon nitride (SiN) photonic devices to achieve polarization-selective filtering and rotation. Our experimental results show that microring resonators incorporating CCPS exhibit transverse magnetic (TM)-pass filtering with a polarization extinction ratio exceeding 25 dB and a low insertion loss of ~ 0.2- 0.4 dB at 1500 -1600 nm. Additionally, under TM-mode input, straight waveguides loaded with CCPS can achieve a significant polarization rotation, with azimuth angle shifts reaching up to 92.9°. Simulations and experimental validation indicate that the primary mechanism behind these effects is the polarization-dependent optical mode overlap, governed by the refractive index profile of the CCPS/SiN hybrid system and the waveguide's geomatrical dimensions. The anisotropic properties of CCPS provide further enhancement but play a secondary role. These results highlight the potential of CCPS-integrated devices for compact, high-performance polarization control in on-chip photonic platforms.

## I. INTRODUCTION

POLARIZATION management is a critical functionality in advanced optical systems, enabling precise control over polarization states[1], [2], [3]. Devices designed for polarization selectivity facilitate the transmission of light with a desired polarization orientation while effectively suppressing orthogonal components[4], [5]. This capability is essential for a wide range of applications, including optical sensing, imaging, display systems, and high-speed optical communication networks[6], [7], [8], [9], [10]. In planar high-index contrast (HIC) integrated photonic circuits, the HIC's guiding materials introduces significant polarization dispersion. As a result, polarization-diversity techniques are indispensable for maintaining enhanced performance, stability, and reliability [11], [12].

Various waveguide polarizers have been developed to achieve polarization selectivity using advanced device architectures and optimized photonic materials. Examples include polarization beam splitters (PBSs), polarization rotators (PRs), and TM-pass or TE-pass polarizers[3], [13], [14]. These devices have been implemented using different methods including geometric asymmetry, subwavelength grating waveguides[15], [16], silicon hybrid plasmonic waveguides[17], and structures based on two-dimensional (2D) materials[18], [19]. Conventionally, waveguides with metal cladding have been widely employed as integrated optical polarizers due to their high polarization extinction ratio (PER) and compact footprint[17]. However, their performance is limited by high insertion loss (IL) and the need for buffer layers to enhance optical operational bandwidth (OBW).

The integration of 2D materials with silicon photonics (SiPh) platform has recently emerged as a promising approach, offering enhanced light-matter interaction, compact device footprints, and compatibility with semiconductor manufacturing processes[20], [21]. Furthermore, 2D materials exhibit unique relevant optical properties that distinguish them from traditional semiconductors, including layer-dependent optical bandgaps, high optical nonlinearity, and pronounced material anisotropy[22], [23]. Various 2D materials have been explored for the realization of optical polarizers, each with distinct advantages and limitations. In such structures, realizing high-performance optical polarization control requires careful selection of 2D materials tailored to specific application requirements [24], [25], [26].

Graphene-based waveguide polarizers have been demonstrated by leveraging the material's anisotropic optical absorption, where TE and TM modes experience different absorption rates. This absorption can be dynamically tuned by adjusting the graphene's chemical potential, for instance, through electrostatic gating, enabling enhanced polarization selectivity[27]. Numerous theoretical and experimental studies have investigated graphene-based waveguide polarizers by optimizing waveguide parameters such as physical dimensions, the presence or absence of an upper cladding, and the substrate refractive index[19], [28]. However, the strong light-matter interaction and inherent absorption of graphene result in increased insertion loss, posing a challenge for practical implementation[27].

Graphene oxide (GO) has also been explored for optical polarizers due to its ease of integration into various photonic

This work was supported by Tamkeen under NYUAD RRC Grant No. CG011. (Corresponding authors: Ghada Dushaq & Mahmoud Rasras). Ghada Dushaq, Solomon Serunjogi, Srinivasa R. Tamalampudi, and Mahmoud Rasras are with the Photonic Research Lab (PRL), Department of Electrical and Computer Engineering, New York University Abu Dhabi, P.O. Box 129188, Abu Dhabi, United Arab Emirates. Fourth Author is also with NYU Tandon School of Engineering, New York University, New York, USA (e-mail: ghd1@nyu.edu; sms10215@nyu.edu; st4212@nyu.edu; mr5098@nyu.edu).

structures, including waveguides and microring resonators[18], [29]. However, the reduction of GO under high optical power must be carefully addressed for practical applications, as it can affect device performance and stability[30].

In our previous work, we demonstrated an electro-optic phase shifter and optical isolation by integrating a new emerging class of 2D ferroionic materials, such as $CuCrP_2S_6$ (CCPS) and and $In_{4/3}P_2Se_6$ [20], [31], [32]. These 2D materials have bandgaps ranging from 1.3 to 3.5 eV and exhibit optical transparency in the short-wave infrared (SWIR) spectrum. Moreover, their high anisotropic refractive index (>3.1 RIU) enhances light-matter interaction, offering advantageous properties not typically found in conventional semiconductors, where high-index materials often exhibit significant absorption in the SWIR region. This work explores the integration of CCPS with silicon nitride (SiN) photonic devices, demonstrating two key functionalities: (1) TM-pass filtering in microring resonators, it achieves a polarization extinction ratio (PER) greater than 25 dB with low insertion loss of 0.2-0.4 dB at 1550 nm, and (2) polarization rotation in straight waveguides, with changes in the azimuth angle exceeding 92° for TM-mode input. These findings highlight the potential of CCPS for compact and efficient on-chip polarization control applications.

The paper is organized as follows: Section II discusses the device design and fabrication. Section III presents the TM-pass filtering in CCPS/SiN microring resonator devices. Polarization rotation is covered in Section IV, followed by the conclusion in Section V.

## II. DEVICE DESIGN AND FABRICATION

### A. SiN Photonic Structures

Figure 1(a) presents an optical image of the fabricated silicon nitride (SiN) photonic chip, which consists of microring resonators and straight waveguide structures. Close-up images of the devices are shown in the inset of Fig. 1(a). Figure 1(b) illustrates the cross-sectional view of the device design. The structure features a 400-nm-thick air-cladded SiN waveguide layer deposited on a 2-µm-thick buried oxide layer. The add-drop microring resonator (MRR) is designed with a waveguide width of 1100 nm and a ring radius of 50 µm. A 120-nm gap between a bus waveguide and the resonator facilitates light coupling, with critical coupling is designed at the through-port output to optimize the extinction ratio.

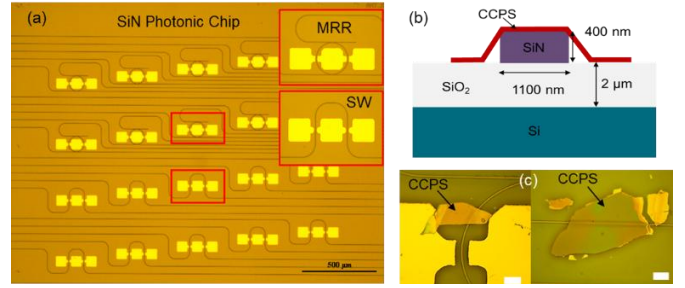

**Fig. 1.** Device design and fabrication (a) Optical image of the fabricated silicon nitride (SiN) photonic chip, with the inset showing close-up images of a straight SiN waveguide and SiN microring resonators. (b) Cross-sectional schematic view of the SiN photonic design parameters. (c) Optical images of CCPS integrated on a microring resonator and a straight waveguide.

### B. CCPS Integration on SiN Photonic Chip

Multilayer CCPS flakes were prepared via mechanical exfoliation from bulk crystals purchased from HQ Graphene. In this work, CCPS flakes with thicknesses ranging from 30 nm to 90 nm were transferred onto SiN microring resonators (MRRs) and straight waveguides using a deterministic dry transfer technique. A detailed description of the transfer methodology is available in prior publications[20], [21], [22], [33].

Figure 1(c) presents optical images of the hybrid SiN/CCPS MRRs and SiN/CCPS straight waveguides. It highlights the precise alignment of the CCPS flakes with the underlying photonic structures. The conformal integration of the CCPS layer ensures efficient evanescent coupling of light from the waveguide to the CCPS layer. The coupling strength is influenced by the thickness of the transferred CCPS flakes, facilitating strong light-matter interaction.

Atomic force microscopy (AFM) was utilized to measure the thickness of the CCPS flakes transferred onto the SiN waveguides, with the results presented in Fig. 2(a). Detailed structural characterization, including energy dispersive X-ray spectroscopy (EDX), X-ray photoelectron spectroscopy (XPS), and transmission electron microscopy (TEM) with diffraction pattern analysis, has been reported in our previous publications[20], [31].

The complex dielectric constant components, $\varepsilon_1$ (real part) and $\varepsilon_2$ (complex part), of the exfoliated CCPS flakes were extracted from high-resolution spectroscopic ellipsometry measurements, as shown in Fig. 2(b). The optical loss tangent, defined as $\tan(\delta)=\varepsilon_2/\varepsilon_1$, was calculated to evaluate material transparency. These values are consistent with the literature and confirm the low optical loss of CCPS flakes in the C-band[20], [34]. The extracted optical parameters of the multilayer CCPS flakes were fed into the mode and beam propagation analyses conducted using the Lumerical Mode Solver.

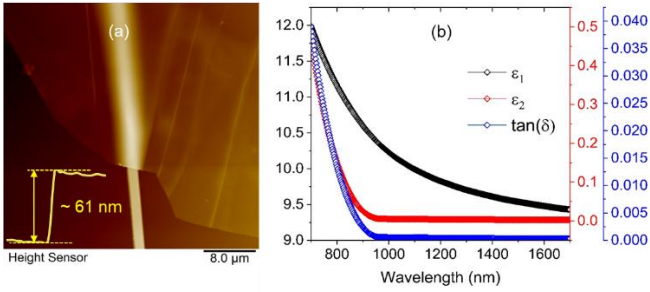

**Fig. 2.** (a) Atomic force microscopy (AFM) image of multilayer CCPS integrated on a SiN waveguide, showing a thickness of approximately 61 nm. (b) Dielectric constants and optical loss tangent extracted from numerical fitting of optical parameters measured via high-resolution spectroscopic ellipsometry.

### III. TM-Pass Filtering

Figure 3(a) illustrates the schematic of the microring resonator integrated with CCPS, demonstrating the configuration of the TM-pass filter. The working principle of the filter is explained using mode simulations, as shown in Fig. 3(b), which depict the mode profiles of the transverse electric (TE) and TM polarizations for both bare and CCPS-loaded SiN waveguides. For the TE mode in the CCPS-loaded structure, the optical mode is shifted into the CCPS layer, leading to signal attenuation due to accumulated optical losses by mode mismatch, which are amplified by the feedback cavity effect of the ring resonator. This leads to the TE mode being effectively filtered out.

In contrast, the TM mode in the bare SiN structure exhibits partial coupling into the substrate. Upon integrating CCPS, the TM mode becomes more confined within the SiN waveguide, significantly enhancing its confinement compared to the bare case. For example, TM mode confinement factor is 40.5% within the bare SiN waveguide. Upon integrating CCPS layers of 30 nm and 60 nm thickness, the TM mode confinement within the SiN waveguide increases to 59.1% and 65.1%, respectively. This demonstrates the role of CCPS in enabling strong TM mode confinement while attenuating the TE mode.

To experimentally quantify the TM-pass filter performance, light was side-coupled into the through-port of the ring resonator using a tunable laser operating in the 1.5–1.6 µm range (C–L optical band). The light was launched via a lensed fiber and collected at the output using another lensed fiber connected to a power meter.

The measured TM transmission spectra for both the bare and CCPS-loaded structures (10 µm interaction length, 60 nm CCPS layer, see Fig.1c) are shown in Fig. 4(a). The results indicate negligible extinction ratio (ER) attenuation in the loaded structure, with certain wavelengths exhibiting higher ER compared to the air-cladded SiN ring, attributed to improved light confinement provided by the CCPS layer.

In contrast, Fig. 4(b) presents the measured TE transmission spectra for the bare and CCPS-loaded ring resonators. The TE mode is almost completely filtered out in the loaded structure, with a measured ER difference of 25 dB between the bare and loaded rings. These experimental results are consistent with the mode simulations, confirming the effectiveness of the TM-pass filter with minimal insertion loss.

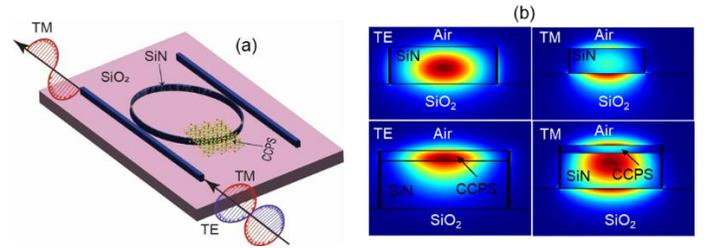

**Fig. 3.** (a) Three-dimensional schematic of the TM-pass filter configuration utilizing CCPS integration. (b) Electric field intensity profiles for TE and TM modes in a bare SiN waveguide (top panel) and a SiN waveguide integrated with a 60-nm-thick CCPS layer (bottom panel) at a wavelength of 1550 nm.

The excess TM-mode loss induced by the integration of CCPS was quantified through a z-transform analysis of the MRR transmission spectra[35], [36]. The round-trip loss of the CCPS-loaded ring was computed and subtracted from that of the bare SiN ring to determine the propagation loss associated with the CCPS material. Results indicate a minimal wavelength-dependent variation in propagation loss across 1500–1550 nm wavelengths, with losses values between 0.2 dB and 0.4 dB for a 10 µm interaction length. This loss is primarily attributed to mode profile mismatches and scattering caused by flake irregularities at the CCPS-SiN interface.

To mitigate these losses, tapering sections between the input/output and hybrid waveguides are proposed to enhance mode overlap and reduce reflection and scattering effects [37], [38]. It is notable that the TM-pass effect is primarily governed by polarization-dependent mode overlap, with the physical dimensions of the SiN waveguide and the refractive index profile of the CCPS/SiN hybrid system playing a critical role in determining the polarization-dependent loss.

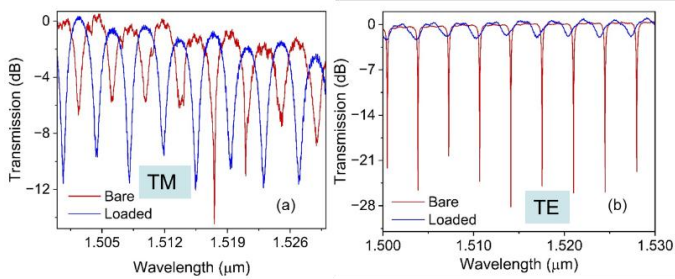

**Fig. 4.** Transmission spectra of bare and CCPS-loaded ring resonators for (a) TM mode and (b) TE mode.

Table 1 presents a comparative evaluation of 2D-material-based optical polarizers with a focus on on-chip integration. The table highlights key experimentally measured performance parameters, including polarization extinction ratio (PER), optical bandwidth (OBW), and insertion loss (IL).

TABLE I
SUMMARY OF ON-CHIP INTEGRATED 2D-MATERIAL-BASED OPTICAL POLARIZERS.

| Material | Platform | Thickness (nm) | PER (dB) | IL (dB) | OBW (µm) | Ref |
|---|---|---|---|---|---|---|
| Gr | Glass WG | - | ~27 | ~9 | 1.23-1.61 | [19] |
| Gr | Polymer WG | 10 | 6 | 9 | - | [28] |
| GO | Doped Silica WG | 2-200 | ~54 | ~7.5 | ~0.63-1.6 | [18] |
| GO | Polymer WG | ~2000 | ~40 | ~6.5 | ~1.53-1.63 | [29] |
| $MoS_2$ | Polymer WG | ~2.5 | ~12.6 | <10 | ~0.65-0.98 | [24] |
| $MoSe_2$ | Polymer WG | 24000 | ~14 | - | ~0.98-1.55 | [25] |
| This work | SiN WG | ~ 60 -80 | ~25 | ~0.2-0.4 | 1.5-1.6 | - |

## IV. POLARIZATION ROTATION

To evaluate the on-chip polarization control capabilities of CCPS, we integrated 120 to 300 µm long CCPS flakes onto 3 mm-long SiN waveguide structures. An on-chip reference MRR test structure with the same waveguide width was employed to facilitate polarization monitoring. Input polarization states were prepared using external waveplates and rotators, ensuring precise coupling into the test ring resonator structure. For instance, TE polarization was calibrated by confirming TE resonances at the output of the ring, and the same procedure was followed for TM polarization.

The calibrated TE or TM input polarization was then coupled into both bare and CCPS-loaded SiN waveguides, and the output polarization states were monitored using the PAX1000 polarimeter. Since the input polarization remained identical for both structures, any change at the output directly corresponded to the impact of CCPS integration. The results demonstrated significant polarization rotation, with the azimuth angle shifting by 77.4° for TE and 92.9° for TM polarization. These observations establish CCPS as a robust platform for polarization control, capable of rotating polarization states with minimal alteration to their ellipticity or handedness.

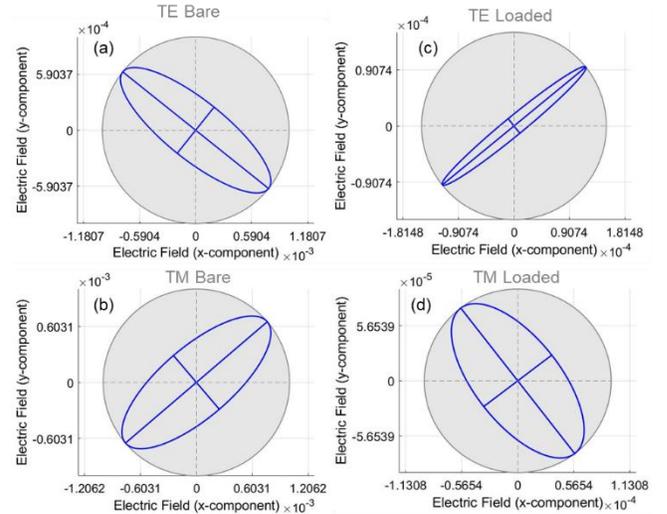

**Fig. 5.** Polarization ellipses recorded using a polarimeter at the output of the SiN waveguide: (a) TE mode in the bare SiN waveguide, (b) TM mode in the bare SiN waveguide, (c) TE mode in the CCPS-loaded SiN waveguide, and (d) TM mode in the CCPS-loaded SiN waveguide.

Figures 5(a) and 5(b) show the output polarization ellipses for the bare SiN waveguide under TE and TM input polarization, respectively. The results confirm a 90° angular separation between the output TE and TM polarizations, verifying the stability and accuracy of the polarization calibration process. In contrast, Figures 5(c) and 5(d) illustrate the output polarization ellipses for the CCPS-loaded SiN waveguide, revealing substantial polarization rotation.

In addition to the polarization ellipses, the Poincaré sphere representation was used to analyze the polarization states for both TE and TM modes in the bare and CCPS-loaded SiN waveguides. Figures 6(a) and 6(b) depict the Poincaré sphere for the TE and TM modes, respectively, in the bare SiN waveguide, while Figs. 6(c) and 6(d) show the corresponding polarization states for the CCPS-loaded waveguides. The Poincaré sphere visualization provides a comprehensive depiction of the polarization rotation, with the azimuth angle shifts for TE and TM modes clearly observable. Specifically, the polarization states for the CCPS-loaded waveguides exhibit significant rotation relative to the bare waveguides, consistent with the polarization ellipse measurements.

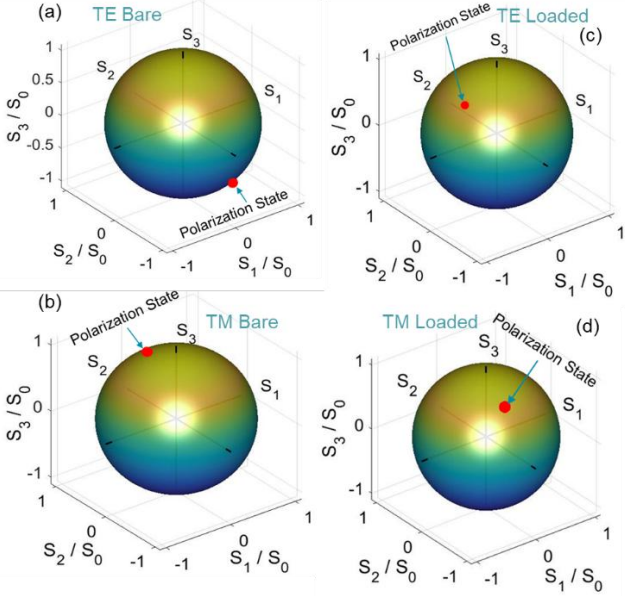

**Fig.6.** Poincaré sphere representation (a) TE mode in the bare SiN waveguide, (b) TM mode in the bare SiN waveguide, (c) TE mode in the CCPS-loaded SiN waveguide, and (d) TM mode in the CCPS-loaded SiN waveguide.

### A. Physical Origins of Polarization Rotation

The polarization rotation in CCPS-loaded SiN waveguides primarily arises at the air/SiN to CCPS/SiN interface, where the abrupt refractive index contrast significantly modifies the boundary conditions of the electromagnetic field. This transition induces a redistribution of the electric field components, effectively altering the polarization state. The mechanisms contributing to this effect include mode hybridization at the interface, differential phase accumulation, and retention of the rotated polarization state.

Mode hybridization occurs due to the sudden change in refractive index at the CCPS-SiN interface, which forces coupling between TE and TM modes. This results in the mixing of transverse and longitudinal electric field components, leading to the formation of new eigenmodes. The strength of this effect is dictated by the interaction length within the CCPS region, where longer propagation enhances mode mixing.

Differential phase accumulation further contributes to polarization rotation as the refractive index difference between TE and TM modes alters their respective phase velocities. The high-index CCPS layer induces a shift in the effective indices of these modes, resulting in an accumulated phase delay that rotates the polarization state. Similar effects are observed in optical fibers and hybrid plasmonic waveguides, where abrupt phase shifts drive polarization transformation[39].

Upon exiting the CCPS region, the rotated polarization state does not revert to its original form but is instead maintained as a stable eigenmode of the SiN waveguide. The phase shift accumulated inside the CCPS region remains embedded in the wave as it propagates, ensuring that the modified polarization state persists beyond the hybrid section.

### B. Simulations Model

To confirm these observations, electromagnetic simulations were performed to analyze the mode evolution, field distribution, and polarization rotation in the CCPS-loaded SiN waveguide. Figure 7(a) presents the schematic of the waveguide-integrated CCPS structure, depicting the calculated effective indices before and after entering the CCPS region. The effective index for TE increases from 1.58 to 1.76, while TM undergoes a smaller shift from 1.47 to 1.58, confirming the role of differential phase accumulation. Figure 7(b) presents the mode profiles for TM-launched light in both the bare and CCPS-loaded waveguides. In the bare waveguide, the TM mode maintains its conventional profile, whereas in the CCPS-loaded waveguide, a clear conversion from TM to TE is observed, providing direct evidence of mode hybridization.

The beam propagation analysis in Figure 7(c) further elucidates the polarization evolution along the x-direction when TM light is launched into the CCPS-loaded region. Eight cross-sectional monitors, placed along the propagation axis, capture the electric field distribution in the zy-plane. The intensity profile of the electric field component Ey undergoes periodic flipping between maxima and minima, which directly indicates energy transfer between TE and TM modes inside the CCPS region. This oscillatory behavior confirms the presence of strong TE-TM mode coupling.

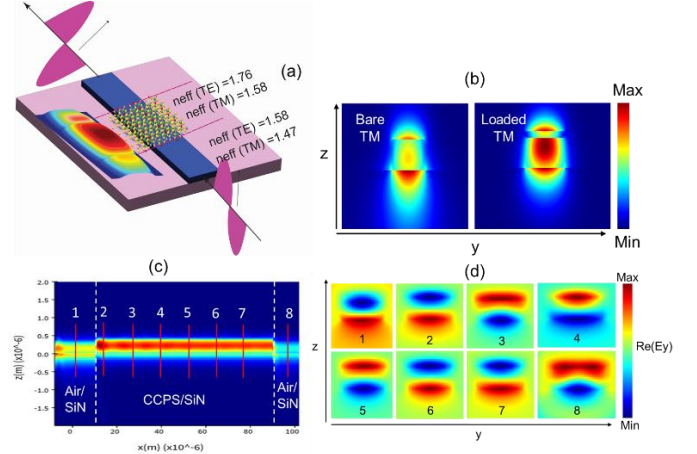

**Fig.7.** (a) Schematic of the waveguide-integrated CCPS structure, depicting the calculated effective indices before and after entering the CCPS region. (b) Simulated TM mode profiles in the bare and CCPS-loaded SiN waveguides, illustrating the effect of CCPS integration on mode confinement and polarization transformation. (c) Beam propagation analysis along the x-direction, showing the evolution of the electric field in the CCPS-loaded waveguide. (d) Cross-sectional field distributions captured at eight monitoring positions, demonstrating the periodic exchange of energy between TE and TM modes, confirming strong TE-TM mode coupling inside the CCPS region.

To quantify the total polarization rotation, a transfer matrix formalism is employed, mathematically representing the cumulative polarization rotation arising within the photonic structure, we define the accumulated phase difference inside CCPS as:

$$\Delta\phi = (\beta_{TE} - \beta_{TM})L \quad (1)$$

Where $\beta_{TE} = \frac{2\pi n_{TE,CCPS}}{\lambda}$ and $\beta_{TM} = \frac{2\pi n_{TM,CCPS}}{\lambda}$

The final polarization state can be approximated in terms of rotation angle $\theta$ (assuming moderate coupling):

$$E_{out} = R(\theta) \cdot E_{in} \quad (2)$$

Where $E_{in} = \begin{bmatrix} E_{TE} \\ E_{TM} \end{bmatrix}$ and the polarization rotation Matrix is

$$R(\theta) = \begin{bmatrix} \cos\theta & -\sin\theta \\ \sin\theta & \cos\theta \end{bmatrix}$$

$\theta \approx \frac{t_{12} t'_{12}}{t_{11} t'_{11}} \sin\Delta\phi$, here $t_{12}$ and $t_{11}$ are the complex coefficient of the first coupling matrix at the transition from air/SiN to CCPS/SiN, while $t'_{12}$ and $t'_{11}$ are the complex coefficient of the coupling matrix on the reverse transition from high-index CCPS back to lower-index SiN, indicating that the polarization rotation angle depends on the mode coupling strength at the interfaces and the phase shift inside CCPS.

## V. Conclusion

This paper demonstrates that the integration of CuCrP$_2$S$_6$ onto silicon nitride photonic devices provides a versatile platform for on-chip polarization control. TM-pass filtering is realized in microring resonators with high polarization extinction ratios, attributed to mode confinement changes caused by CCPS. Polarization rotation experiments on straight waveguides reveal significant azimuth angle shifts due to the abrupt refractive index contrast at the interface and the phase shift inside CCPS. These results underscore the importance of both geometrical mode interaction in enhancing polarization selectivity. The combination of low insertion loss, high performance, and compact device design positions CCPS as a promising material for polarization-sensitive photonic applications.

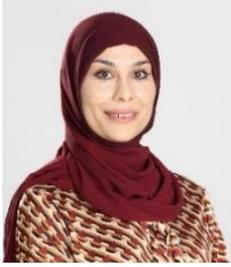
**Ghada Dushaq** (Senior Member, IEEE) is a Research Scientist in the Electrical and Computer Engineering Department at New York University Abu Dhabi, United Arab Emirates. She received her Ph.D. in Interdisciplinary Engineering from the Masdar Institute of Science and Technology, Abu Dhabi, UAE, in 2017, under a cooperative program with the Massachusetts Institute of Technology (MIT) and funded by Mubadala Development Company-Abu Dhabi and GLOBALFOUNDRIES-Singapore. Dr. Dushaq's research focuses on the design and fabrication of advanced optoelectronic devices based on post-silicon materials for applications in data communications, quantum information, and optical sensing. She has received several awards, including the 2018 International Association of Advanced Materials (IAAM) Scientist Medal, the 2020 Falling Walls Lab Prize for her work on "Breaking the Wall of High-Speed Optical Communication" awarded at the World Science Summit in Berlin, the 2021 OWSD-Elsevier Foundation Awards for Women Scientists in Physical Sciences, and the 2021 For Women in Science Middle East Fellowship by L'Oreal-UNESCO. In 2022, Dr. Dushaq was recognized as one of the "10 Innovators Under 35" by MIT Technology Review Arabia. She has published over 80 papers in international peer-reviewed scientific journals and conference proceedings. Dr. Dushaq is a Senior Member of IEEE, and her work has been widely featured in international and local media, including Phys.org, The National, Elsevier, MIT Technology Review, Khaleej Times, and Euronews.

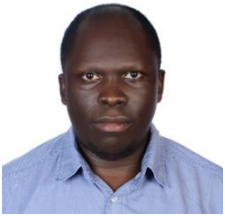
**Solomon Micheal Serunjogi** received the B.Sc. degree in telecommunications from Kyambogo University, Kampala, Uganda, in 2008, and the M.Sc. and Ph.D. degrees from the Masdar Institute of Science and Technology, Abu Dhabi, UAE, in 2013 and 2017, respectively. He is currently with the Electrical and Computer Engineering Department, New York University Abu Dhabi, UAE, involved with the high-speed CMOS transceiver circuits for 400-Gb/s optical links. Prior to joining the New York University, he was with the MTN Uganda, a mobile communication service provider, as a Radio Planning and Optimization Engineer for voice and data services (GSM, WCDMA, WIMAX, and LTE). His research interests include microwave photonics, nonlinear optical signal processing, broadband CMOS circuits, and advanced transmission formats such as PAM-4 and duobinary schemes.

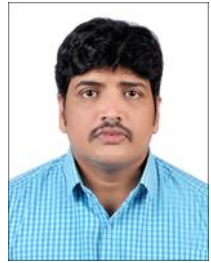
**Srinivasa R. Tamalampudi** received the M.S. degree in microsystems engineering from Hochschule Furtwangen, Germany, in 2009, and the Ph.D. degree in physics from Academia Sinica, Taiwan International Graduate Program (TIGP), Taiwan, in 2015. From 2016 to 2017, he was a Research Associate with the University of Cambridge, Cambridge, U.K. From 2017 to 2020, he was a Postdoctoral Fellow with the Khalifa University of Science and Technology, Abu Dhabi, UAE. He is currently a Research Associate with the New York University of Abu Dhabi, Abu Dhabi, UAE. His current research interests include heterogeneous integration of 2D materials onto Si and SiN photonic chips for optoelectronic applications, such as photo-detectors, modulators and light emitting devices.

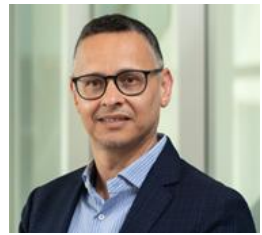
**Mahmoud Rasras** (Senior Member, IEEE) is a Professor of Electrical Engineering at New York University Abu Dhabi (NYUAD). He received his Ph.D. in Physics from the Catholic University of Leuven, Belgium, where he conducted research at IMEC, focusing on the development of spectroscopic photon-emission microscopy techniques for failure analysis of semiconductor (CMOS) devices. Dr. Rasras has more than 11 years of industrial research experience. He was a Member of the Technical Staff at Bell Labs, Alcatel-Lucent, NJ, USA. His work covered a broad range of activities, particularly the development of novel photonic components for next-generation optical and microwave photonic networks. At NYUAD, his research group has developed designs for silicon photonics circuits, 2D-based photodetectors and isolators, and high-speed germanium photodetectors. He served as an Associate Editor for *Optics Express* and has authored and co-authored more than 200 peer-reviewed papers. Before joining NYUAD, he was a faculty member at Masdar Institute and the Director of the SRC/GlobalFoundries Center of Excellence for Integrated Photonics. Dr. Rasras is a Senior IEEE Member and a member of the Mohammed bin Rashid Academy of Scientists (MBRAS).